\documentclass{article}

\PassOptionsToPackage{numbers, compress}{natbib}
\usepackage[preprint]{neurips_2023}

\usepackage{threeparttable}
\usepackage[utf8]{inputenc} 
\usepackage[T1]{fontenc}    
\usepackage{hyperref}       
\usepackage{url}            
\usepackage{booktabs}       
\usepackage{amsfonts}       
\usepackage{nicefrac}       
\usepackage{microtype}      
\usepackage{xcolor}         
\usepackage{amsmath}
\usepackage{amssymb}
\usepackage{mathtools}
\usepackage{amsthm}
\usepackage{multirow}
\usepackage{wrapfig}
\usepackage[capitalize,noabbrev]{cleveref}
\usepackage{graphicx}
\usepackage{subfigure} 
\usepackage{float}
\usepackage{algorithm} 
\usepackage{algorithmic}
\usepackage{textcomp} 
\usepackage{listings} 
\usepackage{enumitem}
\usepackage{utfsym}
\usepackage{times}
\usepackage{latexsym}
\usepackage{fancyvrb}
\usepackage{CJKutf8}

\newcommand{\commentout}[1]{}

\title{FireRedASR: Open-Source Industrial-Grade Mandarin Speech Recognition Models\\from Encoder-Decoder to LLM Integration}

\author{Kai-Tuo Xu, Feng-Long Xie, Xu Tang, Yao Hu \\
Xiaohongshu Inc.
}

\begin{document}
\maketitle
\begin{abstract}
We present FireRedASR, a family of large-scale automatic speech recognition (ASR) models for Mandarin, designed to meet diverse requirements in superior performance and optimal efficiency across various applications. FireRedASR comprises two variants:

\textbf{FireRedASR-LLM}: Designed to achieve state-of-the-art (SOTA) performance and to enable seamless end-to-end speech interaction. It adopts an Encoder-Adapter-LLM framework leveraging large language model (LLM) capabilities. On public Mandarin benchmarks, FireRedASR-LLM (8.3B parameters) achieves an average Character Error Rate (CER) of 3.05\%, surpassing the latest SOTA of 3.33\% with an 8.4\% relative CER reduction (CERR). It demonstrates superior generalization capability over industrial-grade baselines, achieving 24\%-40\% CERR in multi-source Mandarin ASR scenarios such as video, live, and intelligent assistant.

\textbf{FireRedASR-AED}: Designed to balance high performance and computational efficiency and to serve as an effective speech representation module in LLM-based speech models. It utilizes an Attention-based Encoder-Decoder (AED) architecture. On public Mandarin benchmarks, FireRedASR-AED (1.1B parameters) achieves an average CER of 3.18\%, slightly worse than FireRedASR-LLM but still outperforming the latest SOTA model with over 12B parameters. It offers a more compact size, making it suitable for resource-constrained applications.

Moreover, both models exhibit competitive results on Chinese dialects and English speech benchmarks and excel in singing lyrics recognition. To advance research in speech processing, we release our models and inference code at \url{https://github.com/FireRedTeam/FireRedASR}.

\end{abstract}



\section{Introduction}
Automatic Speech Recognition (ASR) has evolved rapidly in recent years, becoming an essential component in intelligent voice interaction and multimedia content understanding. 
Recent advances in ASR have led to several large-scale models, such as Whisper \cite{radford2023robust}, Qwen-Audio \cite{chu2023qwen,chu2024qwen2}, SenseVoice \cite{an2024funaudiollm}, and Seed-ASR \cite{seedasr2024}, showing a paradigm shift from end-to-end models with millions of parameters \cite{li2022recent,prabhavalkar2023end} to larger-scale models \cite{radford2023robust,an2024funaudiollm,zhang2023google,song2024touchasp} and the integration of pre-trained text LLMs \cite{chu2023qwen,chu2024qwen2,seedasr2024,wu2023decoder,rubenstein2023audiopalm,li2023prompting,wang2023slm,pan2023cosmic,yu2024connecting,chen2024salm,lakomkin2024end,geng2024unveiling,ma2024embarrassingly}.

Despite their impressive capabilities and larger model sizes, they face significant limitations in practical applications. Some models prioritize multilingual and multitask capabilities, resulting in suboptimal performance for specific languages like Mandarin. Others, despite showing promising results, are limited by their closed-source nature, restricting community-driven improvements and academic research. The growing demands for modern speech interaction systems, highlighted by GPT-4o \cite{openai2024gpt4o,hurst2024gpt}, underscore the need for open-source, high-performance Mandarin ASR solutions.

To address these limitations, in this technical report, we introduce FireRedASR, a family of large-scale models for Mandarin ASR. To address varying needs in performance and efficiency across a wide range of application scenarios, FireRedASR consists of two variants: FireRedASR-LLM  
 and FireRedASR-AED. FireRedASR-LLM utilizes an innovative Encoder-Adapter-LLM framework \cite{seedasr2024,wu2023decoder,geng2024unveiling,ma2024embarrassingly}, comprising 8.3B parameters to push the boundary of recognition accuracy. This model is particularly well-suited for scenarios where precision is paramount and computational resources are not a primary constraint. FireRedASR-AED, on the other hand, is designed to balance superior performance and optimal efficiency. It employs an Attention-based Encoder-Decoder (AED) architecture \cite{bahdanau2016end,chan2016listen} with up to 1.1B parameters. Beyond its standalone use, FireRedASR-AED also functions as a crucial speech representation component within larger LLM-based speech frameworks.

Key contributions of our work include:
\begin{itemize}
\item \textbf{High-Accuracy Models with Efficiency}: On public Mandarin benchmarks, FireRedASR-LLM achieves an average Character Error Rate (CER) of 3.05\%, surpassing the previous state-of-the-art (Seed-ASR) of 3.33\% with an 8.4\% relative reduction. Meanwhile, FireRedASR-AED attains a  CER of 3.18\%, outperforming Seed-ASR (over 12B parameters) with significantly fewer parameters. These results highlight the ability of our models to achieve superior accuracy while maintaining efficiency.
\item \textbf{Robust Real-World Performance}: In diverse practical scenarios, including short videos, live streaming, auto-captioning, voice input, and intelligent assistants, our models demonstrate exceptional capabilities, achieving 24\%-40\% relative CER reduction (CERR) compared to popular open-source baseline and leading commercial solutions.
\item \textbf{Versatile Recognition Capabilities}: Both variants demonstrate remarkable versatility beyond standard Mandarin ASR, showing competitive results on Chinese dialects and English speech benchmarks. Notably, they achieve 50\%-67\% CERR in singing lyrics recognition compared to industrial-grade baselines.
\item \textbf{Comprehensive Open-Source Release}: We contribute to the research community by releasing our model family, including pre-trained weights and efficient inference code. This open-source release aims to accelerate research progress in speech processing and enable broader applications in modern end-to-end speech interaction systems.
\end{itemize}

The remainder of this report is organized as follows: Section 2 describes the architectures of FireRedASR-AED and FireRedASR-LLM, along with training data and optimization strategies. Section 3 presents comprehensive evaluation results across various benchmarks and practical scenarios compared to recently released large-scale ASR models. Section 4 discusses the key factors contributing to our superior performance. Section 5 concludes the report.

\section{FireRedASR}
In this section, we present the architectural details and methodologies for our two ASR models: FireRedASR-AED and FireRedASR-LLM. FireRedASR-AED follows the conventional Attention-based Encoder-Decoder architecture, whereas FireRedASR-LLM is built on the Encoder-Adapter-LLM architecture that leverages the power of LLM for ASR. Both models share similar input feature processing and acoustic encoding strategies but differ in their approaches to token sequence modeling.

\begin{figure*}[!ht]
\centering
\includegraphics[width=\linewidth]{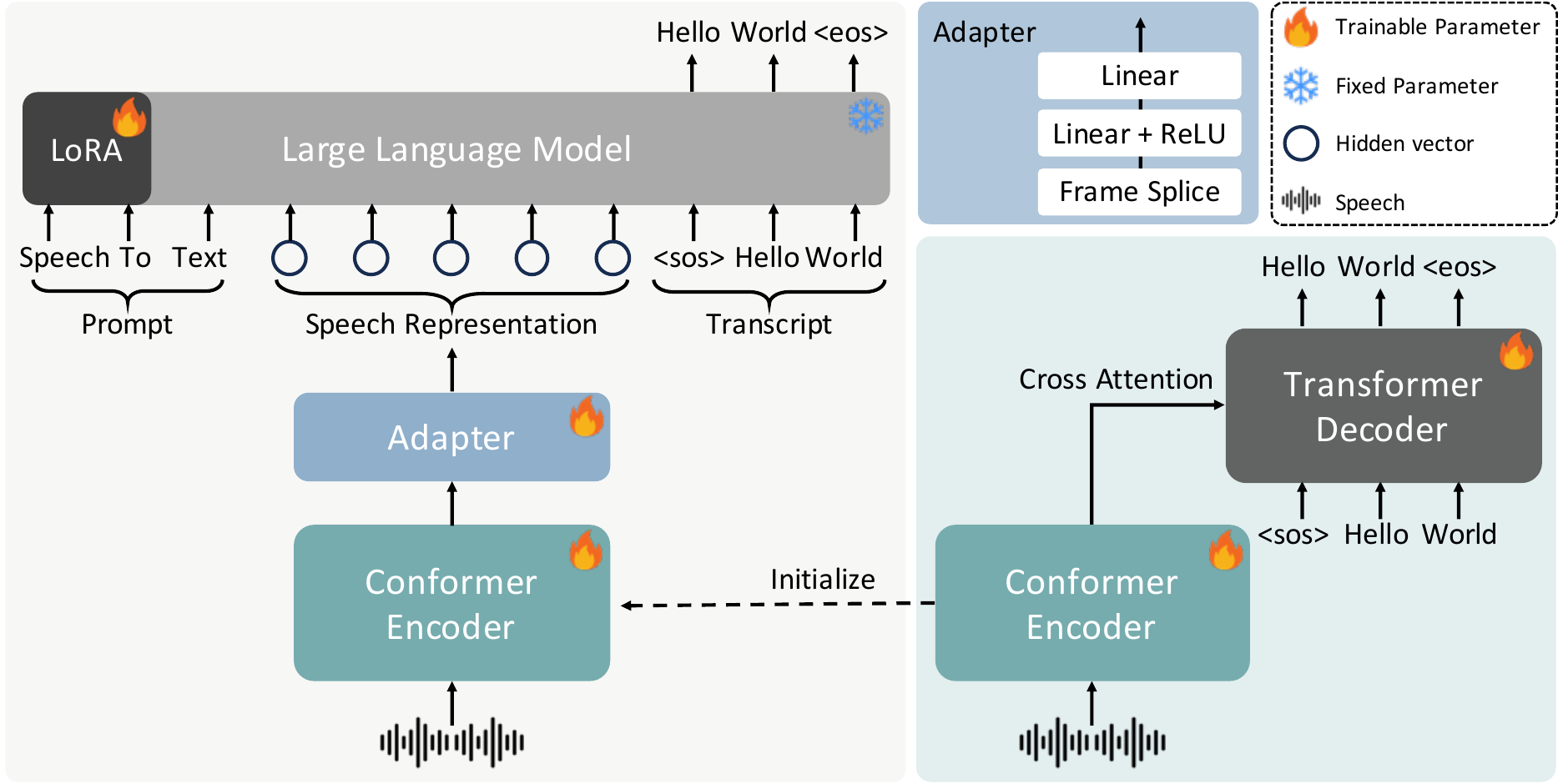}
\caption{Architecture of FireRedASR-LLM (left), FireRedASR-AED (bottom right), and Adapter.}
\label{img:llm_asr}
\end{figure*}

\subsection{FireRedASR-AED: Attention-based Encoder-Decoder ASR model}
FireRedASR-AED adopts an end-to-end architecture that combines a Conformer-based Encoder (Enc) with a Transformer-based Decoder (Dec)\cite{gulati2020conformer,vaswani2017attention}. This design choice leverages both the ability of Conformer to model local and global dependencies in speech features and the effectiveness of Transformer in sequence transduction. The overall architecture of FireRedASR-AED is illustrated in Figure \ref{img:llm_asr} (bottom right).

\noindent\textbf{Training Data}: The training corpus consists of approximately 70,000 hours of audio data, predominantly high-quality Mandarin Chinese speech. Unlike weakly-labeled datasets used in Whisper, the majority of our data was manually transcribed by professional annotators, ensuring high transcription accuracy and reliability. The dataset also incorporates approximately 11,000 hours of English speech data to enhance English ASR capabilities.

\noindent\textbf{Input Features}: The input features are 80-dimensional log Mel filterbank (Fbank) extracted from 25ms windows with 10ms frame shifts, followed by global mean and variance normalization.

\noindent\textbf{Encoder Structure}: The encoder consists of two main components: a subsampling module and a stack of Conformer blocks. The subsampling module employs two sequential convolutional layers, each with a stride of 2 and a kernel size of 3, followed by ReLU activation functions. This configuration reduces the temporal resolution from 10ms to 40ms per frame, effectively managing computational complexity while preserving essential acoustic information.
The subsampled features are then processed by a stack of Conformer blocks. Each Conformer block consists of four primary components: two Macaron-style feedforward modules positioned at the beginning and end of the block, a multi-head self-attention module incorporating relative positional encoding \cite{dai2019transformer}, and a convolution module equipped with gated linear unit (GLU) and layer normalization. The kernel size for all 1-D depthwise convolution is set to 33. This structure enables effective modeling of both local and global dependencies in the speech signal, while maintaining computational efficiency.

\noindent\textbf{Decoder Structure}: The decoder follows a standard Transformer architecture with several key design choices. It adopts fixed sinusoidal positional encodings and employs weight tying between input and output token embeddings to reduce model complexity. Each Transformer block consists of three primary components: a multi-head self-attention module, a multi-head cross-attention module, and a position-wise feedforward module, all utilizing pre-norm residual units to enhance training stability and gradient flow.

\noindent\textbf{Tokenization}: We employ a mixed tokenization strategy: Chinese characters for Chinese text and token-level byte-pair encoding (BPE) \cite{sennrich2015neural} for English text. The total vocabulary size is 7,832, comprising 1,000 English BPE tokens, 6,827 Chinese characters, and 5 special tokens.

We investigated various sizes of FireRedASR-AED, with detailed architectural configurations presented in Table \ref{tab:model_config}, where \#Params denotes the number of parameters. Unless otherwise specified, FireRedASR-AED refers to FireRedASR-AED-L.




\begin{table} [!ht]
    \caption{Architecture details of FireRedASR-AED and FireRedASR-LLM.}
    \label{tab:model_config}
    \centering
    \begin{tabular}{l|rrrr}
        \toprule
        \textbf{Model Size} & \textbf{XS} & \textbf{S} & \textbf{M} & \textbf{L} \\ 
        \midrule
        \multicolumn{5}{c}{\textbf{FireRedASR-AED}} \\
        \midrule
        Width ($d_{model}$) & 512  & 768  & 1024 & 1280  \\
        \#Layers (Enc/Dec) & 12/12 & 16/16 & 16/16 & 16/16 \\
        \#Params (Total)     & 140M & 413M & 732M & \textbf{1.1B} \\
        \midrule
        \multicolumn{5}{c}{\textbf{FireRedASR-LLM}} \\
        \midrule
        \#Params (Encoder)  & 86M  & 256M & 455M & 710M  \\
        \#Params (Adapter)  & 17M  & 18M & 20M & 22M  \\
        \#Params (Total)      & 7.7B & 7.9B & 8.1B & \textbf{8.3B} \\
        \bottomrule
    \end{tabular}
\end{table}
         
\subsection{FireRedASR-LLM: Encoder-Adapter-LLM-based ASR model}
FireRedASR-LLM is also an end-to-end ASR model but designed to integrate robust speech processing capabilities of FireRedASR-AED with the superior language capabilities of LLM. It comprises three core components: a Conformer-based audio Encoder, a lightweight audio-text alignment Adapter and a pre-trained text-based LLM, forming what we term the Encoder-Adapter-LLM architecture. The overall architecture of FireRedASR-LLM is illustrated in Figure \ref{img:llm_asr} (left).

\noindent\textbf{Input Features and Encoder}: FireRedASR-LLM employs the same training data, input features and processing methods as FireRedASR-AED. The encoder of FireRedASR-LLM is initialized with pre-trained weights from Encoder of FireRedASR-AED. This encoder generates continuous representations that encapsulate both acoustic and semantic characteristics of the input speech. 

\noindent\textbf{Adapter Structure and Functionality}: To seamlessly integrate the audio encoder with the text-based LLM, an adapter network is employed. This adapter transforms the output of encoder into the semantic space of the LLM, enabling the LLM to accurately recognize the corresponding text content from the input speech. The adapter consists of a simple but effective Linear-ReLU-Linear network, which projects the output dimension of encoder to match the input embedding dimension of the LLM.
Even after temporal subsampling from 10ms to 40ms, the output of the encoder remains too lengthy for the LLM to process efficiently. Therefore, we incorporate an additional frame splicing operation at the beginning of the adapter. This operation further reduces the temporal resolution from 40ms to 80ms per frame, thereby decreasing sequence length and improving the computational efficiency for the LLM.

\noindent\textbf{LLM Initialization and Processing}: The LLM component of FireRedASR-LLM is initialized with pre-trained weights from Qwen2-7B-Instruct \cite{qwen2}, a notable open-source LLM.
During training, the input of FireRedASR-LLM consists of a triplet: (prompt, speech, transcript). The encoder and adapter produces a speech embedding ${E}_{S}$, while the prompt and transcript are tokenized and embedded by the LLM into prompt embedding ${E}_{P}$ and transcript embedding ${E}_{T}$. These embeddings are concatenated as $({E}_{P}, {E}_{S}, {E}_{T})$ and processed by the subsequent layers of LLM.
During inference, the input is reduced to $({E}_{P},{E}_{S})$, enabling the LLM to execute next-token-prediction and generate recognized text from speech.

\noindent\textbf{Training Strategy}: We employ a carefully designed training strategy that balances adaptation and preservation of pre-trained capabilities: the encoder and adapter are fully trainable, while the majority of LLM parameters remain fixed. We incorporate trainable LLM Low-Rank Adaptation (LoRA) \cite{hu2021lora} to efficiently fine-tune the LLM. This strategy ensures that the encoder and adapter are adequately trained to map speech features into the semantic space of LLM, while preserving its pre-trained capabilities.
The training objective is based on cross-entropy loss, with the loss computed only over the transcript portion of the input, ignoring the prompt and speech embeddings.

We investigated various sizes of FireRedASR-LLM, with detailed architectural configurations presented in Table \ref{tab:model_config}. Unless otherwise specified, FireRedASR-LLM refers to FireRedASR-LLM-L.

\section{Evaluation}
In this section, we conduct a comprehensive evaluation of FireRedASR-LLM and FireRedASR-AED models, with a primary focus on their performance in Mandarin speech recognition. The evaluation is structured into three parts to systematically assess the capabilities and generalization abilities of the models.

First, we benchmark our models using several public Mandarin test sets to establish baseline performance under standardized conditions. Second, we evaluate their performance on diverse multi-source Mandarin speech test sets to validate their robustness in real-world scenarios. Additionally, we assess the models' effectiveness in singing lyrics recognition, crucial for specific industrial applications. Third, we evaluate the models' performance on Chinese dialects and English speech recognition to demonstrate their potential for broader applications beyond standard Mandarin.

\noindent\textbf{Metrics}: We use Character Error Rate (CER) for evaluating Chinese speech and singing lyrics recognition, and Word Error Rate (WER) for English.

\subsection{Evaluation on Public Mandarin ASR Benchmarks}
We benchmark FireRedASR-LLM and FireRedASR-AED compared to several recently released large-scale ASR models, including Seed-ASR \cite{seedasr2024}, SenseVoice-L \cite{an2024funaudiollm}, Qwen-Audio \cite{chu2023qwen}, Paraformer-Large \cite{gao2022paraformer}, and Whisper-Large-v3 \cite{radford2023robust}. The evaluation is conducted on four widely-used public Chinese Mandarin ASR test sets: 1) AISHELL-1 \cite{bu2017aishell} test set (aishell1); 2) AISHELL-2 \cite{du2018aishell} iOS version test set (aishell2); 3) WenetSpeech \cite{zhang2022wenetspeech} Internet domain test set (ws\_net); 4) WenetSpeech meeting domain test set (ws\_meeting). The results for the comparative models are sourced from their respective publications, with Whisper-Large-v3 results taken from the SenseVoice-L \cite{an2024funaudiollm} and WenetSpeech results of Qwen-Audio derived from the Seed-ASR \cite{seedasr2024}.

As illustrated in Table \ref{tab:test_public}, both FireRedASR-LLM and FireRedASR-AED outperform Seed-ASR. Notably, FireRedASR-LLM achieves an 8.4\% relative CER reduction (CERR) compared to Seed-ASR when averaged across all four test sets (Average-4). Seed-ASR, a state-of-the-art large ASR models but not open-source, has been trained with 7.7 million hours in its self-supervised learning stage and 0.562 million hours in its supervised fine-tuning stage, with nearly 2B parameters in its encoder and over 10B parameters in its LLM \cite{seedasr2024}. In contrast, FireRedASR-AED contains only 1.1B parameters and FireRedASR-LLM includes 8.3B parameters, highlighting the effectiveness of our models' architecture, training strategies and datasets. When compared to other models, most of which are open-source, FireRedASR-AED achieves a 29\%-68\% CERR with fewer parameters than Whisper-Large-v3, SenseVoice-L, and Qwen-Audio.

\textbf{Observation of Scaling Law}: Recent studies in LLMs have demonstrated that model performance typically improves with increased model size, known as the scaling law \cite{kaplan2020scaling}. As shown in \ref{tab:test_size}, we investigate the scaling behavior of our models with different model sizes, as detailed in Table \ref{tab:model_config}. For FireRedASR-AED, we scale the model sizes progressively from 140M, 413M, 732M to 1.1B parameters. The performance consistently improves with increased model size, achieving CERRs of 6.1\%, 5.3\%, and 5.6\% when scaling from XS to S, S to M, and M to L configurations respectively. For FireRedASR-LLM, we focus on scaling the encoder while keeping the LLM backbone unchanged. The encoder size increases from 86M to 710M parameters, with minimal changes in adapter parameters (17M to 22M). This exhibits similar scaling patterns and leads to consistent performance improvements, with an overall 7.3\% CERR from XS (3.29\%) to L (3.05\%) configuration. These results demonstrate the effectiveness of our scaling strategies and suggest the potential for further improvements with larger model capacities.

\begin{table}[!ht]
    \caption{Comparison of Character Error Rate (CER\%) for FireRedASR-LLM, FireRedASR-AED and other released large ASR models on four public Mandarin ASR test sets.}
    \label{tab:test_public}
    \centering
    \begin{tabular}{l|c|cccc|c}
        \toprule
        Model & \#Params & aishell1 & aishell2 & ws\_net & ws\_meeting & Average-4\footnotemark[1] \\ 
        \midrule
        FireRedASR-LLM  & 8.3B & 0.76 & 2.15 & 4.60 & 4.67 & \textbf{3.05} \\
        FireRedASR-AED  & 1.1B & 0.55 & 2.52 & 4.88 & 4.76 & 3.18 \\
        \midrule
        Seed-ASR     & 12B+ & 0.68 & 2.27 & 4.66 & 5.69 & 3.33 \\
        Qwen-Audio   & 8.4B & 1.30 & 3.10 & 9.50 & 10.87 & 6.19\\
        SenseVoice-L & 1.6B & 2.09 & 3.04 & 6.01 & 6.73 & 4.47\\
        Whisper-Large-v3 & 1.6B & 5.14 & 4.96 & 10.48 & 18.87 & 9.86 \\
        Paraformer-Large & 0.2B & 1.68 & 2.85 & 6.74 & 6.97 & 4.56 \\
        \bottomrule
    \end{tabular}
\end{table}
\footnotetext[1]{Seed-ASR reports an average CER across six public Mandarin sets (Average-6), including the four sets discussed here plus AISHELL-2 Android and Mic versions. We focus on Average-4 as the latter two differ from the iOS version only in recording devices and the iOS version is more commonly evaluated in the literature. For direct comparison, our FireRedASR-LLM achieves 2.86\% Average-6 CER, outperforming Seed-ASR's 2.98\%.}

\begin{table}
    \caption{Comparison of average CER for FireRedASR-LLM and FireRedASR-AED with different model size on four public Mandarin ASR test sets.}
    \label{tab:test_size}
    \centering
    \begin{tabular}{l|cccc}
        \toprule
        Model Size  & XS  & S & M & L \\
        \midrule
        FireRedASR-LLM   & 3.29 & 3.23 & 3.19 & 3.05 \\
        FireRedASR-AED   & 3.79 & 3.56 & 3.37 & 3.18 \\
        \bottomrule
    \end{tabular}
\end{table}

\subsection{Evaluation on Multi-source Mandarin Speech and Singing Benchmarks}
To comprehensively evaluate the capabilities of FireRedASR-LLM and FireRedASR-AED, we conduct extensive testing on both multi-source Mandarin speech recognition and singing lyrics recognition. The speech test sets are carefully curated from five diverse scenarios: short videos, live streaming, auto-captioning, voice input, and intelligent assistant, ensuring broad coverage of real-world applications. We calculate the average CER across these scenarios to ensure robust evaluation. Additionally, we construct a singing lyrics test set from short videos to assess singing lyrics recognition performance, which is a critical requirement for various practical applications.

For comparative analysis, we select two categories of baseline systems: 1) Paraformer-Large, a widely adopted open-source model in the Mandarin speech processing community, and 2) commercial ASR services from a leading Mandarin ASR provider (denoted by ProviderA) in the industry, including both their base (ProviderA-Base) and large (ProviderA-Large) versions.

As shown in Table \ref{tab:test_internal}, in the speech recognition task, FireRedASR-LLM achieves the best performance with a CER of 3.48\%, followed closely by FireRedASR-AED with 3.74\%. Both models significantly outperform the commercial and open-source baselines, with FireRedASR-LLM showing a 23.7\% relative improvement over ProviderA-Large (CER 4.56\%) and a 38.6\% relative improvement over Paraformer-Large (CER 5.80\%).

In the singing lyrics recognition task, the performance gap becomes even more pronounced. FireRedASR-LLM maintains superior performance with a CER of 7.05\%, while ProviderA-Large and Paraformer-Large show substantially higher CER of 14.16\% and 21.19\% respectively, corresponding to CERR of 50.2\% and 66.7\%. This remarkable improvement in singing lyrics recognition demonstrates the robust capability of our models in handling challenging acoustic conditions and varying vocal styles.

Notably, FireRedASR-AED also maintains significant advantages over other baseline systems in both speech and singing lyrics recognition tasks. These results convincingly demonstrate that both FireRedASR-AED and FireRedASR-LLM have achieved superior industrial-grade performance, with particular strength in handling diverse acoustic conditions and specialized tasks like singing lyrics recognition.

\begin{table}[!ht]
    \caption{Comparison of CER and relative CER reduction (CERR) for FireRedASR-LLM, FireRedASR-AED and baseline ASR models on multi-source Mandarin speech and singing test sets. CERR values are computed relative to FireRedASR-LLM performance.}
    \label{tab:test_internal}
    \centering
    \begin{tabular}{l|cc|cc}
        \toprule
        \multirow{2}{*}{Model} & \multicolumn{2}{c|}{Speech} & \multicolumn{2}{c}{Singing} \\
        \cmidrule{2-3} \cmidrule{4-5}
        &CER(\%) & CERR & CER(\%) & CERR \\
        \midrule
        FireRedASR-LLM    & \textbf{3.48}  & 0.0\% & \textbf{7.05} & 0.0\% \\
        FireRedASR-AED    & 3.74 & 7.0\% & 7.51 & 6.1\% \\
        \midrule
        ProviderA-Large    & 4.56 & 23.7\% & 14.16 & 50.2\% \\
        ProviderA-Base     & 5.67 & 38.6\% & 21.37 & 67.0\% \\
        Paraformer-Large  & 5.80 & 40.0\% & 21.19 & 66.7\% \\
        \bottomrule
    \end{tabular}
\end{table}

\subsection{Evaluation on Public Chinese Dialect and English ASR Benchmarks}
FireRedASR-LLM and FireRedASR-AED exhibit strong generalization capabilities, achieving impressive results on Chinese dialect and English speech recognition despite being primarily designed for Mandarin ASR. To demonstrate the models' effectiveness beyond standard Mandarin, we evaluate their performance on several widely-adopted public benchmarks. To the best of our knowledge, we compare our models with the previous SOTA open-source models on these respective test sets. 

For Chinese dialect speech recognition, we evaluate our models on the KeSpeech \cite{tang2021kespeech} test set. According to the recently released report \cite{li2024baichuan}, existing models including Baichuan-omni, Qwen2-Audio-Instruct, and Whisper-Large-v3 (with parameter sizes of 7B+, 7B+, and 1.5B respectively) achieve average CERs of 6.7\%, 9.9\%, and 44\% on KeSpeech. As shown in Table \ref{tab:test_other}, both FireRedASR-LLM and FireRedASR-AED significantly outperform these models, achieving CERs of 3.56\% and 4.48\% respectively.

For English speech recognition, we evaluate our models on the widely-adopted LibriSpeech \cite{panayotov2015librispeech} test sets (test-clean and test-other). Whisper-Large-v3, a popular open-source multilingual ASR model trained on 5 million hours of audio data, achieves WERs of 1.82\% and 3.50\% on test-clean and test-other respectively, as reported in \cite{an2024funaudiollm}. Our models demonstrate competitive performance: FireRedASR-LLM achieves WERs of 1.73\% and 3.67\%, while FireRedASR-AED achieves WERs of 1.93\% and 4.44\% on the respective test sets.

\begin{table}[!ht]
    \caption{Comparison of ASR performance on Chinese dialect (KeSpeech) and English (LibriSpeech) test sets. Results are reported in CER(\%) for KeSpeech and WER(\%) for Librispeech. Previous SOTA open-source results are from \cite{li2024baichuan,radford2023robust,an2024funaudiollm}.}
    \label{tab:test_other}
    \centering
    \begin{tabular}{l|ccc}
        \toprule
        Test Set       & KeSpeech & LibriSpeech test-clean & LibriSpeech test-other \\
        \midrule
        FireRedASR-LLM &   3.56 & 1.73 & 3.67\\
        FireRedASR-AED &   4.48 & 1.93 & 4.44 \\
        Previous SOTA Results & 6.70 & 1.82 & 3.50  \\
        \bottomrule
    \end{tabular}
\end{table}

\section{Discussion}
In this section, we explore the reasons why our FireRedASR models outperform competing models. We attribute the superior performance to the following three factors:

\textbf{High-Quality and Diverse Training Data}: Our training corpus consists predominantly of professionally transcribed audio collected from real-world scenarios, which provides significantly more valuable training signals than traditional reading-style recordings in controlled environments. The dataset encompasses extensive variations in acoustic conditions, speakers, accents, and content domains, totaling tens of thousands of hours. Such diversity and scale enable our models to learn robust speech representations and linguistic patterns, leading to strong generalization. Our empirical studies demonstrate that one thousand hours of high-quality, human-labeled data yields better results than ten thousand hours of weakly-labeled data (e.g., from video captions, OCR results, or ensemble ASR outputs), explaining our advantage over Whisper-like models. Moreover, the inclusion of singing data in our corpus contributes to our significant performance improvements over baseline models in handling musical content.

\textbf{Optimized Training Strategy}: When scaling FireRedASR-AED from 140M to 1.1B parameters, we identified regularization and learning rate as critical factors affecting model convergence. We developed a \textbf{Progressive Regularization Training} strategy: initially training without regularization techniques (dropout and SpecAugment \cite{park2019specaugment}) to achieve rapid convergence, then gradually introducing stronger regularization as overfitting tendencies emerge. This method enabled successful training of the FireRedASR-AED 1.1B, demonstrating superior outcomes. The strategy proved beneficial for smaller models with 732M, 413M, and 140M parameters as well. Furthermore, larger models benefit from reduced learning rates, making it crucial to adjust this parameter for optimal performance.

\textbf{Efficient ASR Framework}: Our architectural choices were informed by extensive experimentation and prior work. While our previous Two-pass Transducer-based model \cite{graves2012sequence,sainath2019two} achieved reasonable performance across various ASR models with millions of parameters, it exhibited scaling limitations and high sensitivity to hyperparameters, with the Prediction Network component prone to overfitting. The Transducer approach also imposed significant memory overhead compared to the cross-entropy loss used in FireRedASR. Drawing inspiration from recent advances like Whisper while addressing these limitations, we adopted an attention-based encoder-decoder architecture enhanced with our implementations of Conformer and Transformer. Furthermore, we incorporated a simple yet effective adapter design inspired by recent works \cite{chu2023qwen,chu2024qwen2,seedasr2024,wu2023decoder,rubenstein2023audiopalm,li2023prompting,wang2023slm,pan2023cosmic,yu2024connecting,chen2024salm,lakomkin2024end,geng2024unveiling,ma2024embarrassingly}, facilitating efficient model adaptation and research iteration.

\section{Conclusion}

We have presented FireRedASR-LLM and FireRedASR-AED, two high-performance ASR models optimized for Mandarin. Through comprehensive evaluations, we demonstrate that their architectures, training strategies, and high-quality datasets can achieve state-of-the-art performance while maintaining computational efficiency. FireRedASR-AED proves that attention-based encoder-decoder architectures remain highly competitive, while FireRedASR-LLM, leveraging the Encoder-Adapter-LLM framework, showcases the potential of integrating LLM capabilities into ASR systems.
Our extensive evaluation results confirm the strong performance of both models across multiple dimensions: achieving state-of-the-art results on public Mandarin benchmarks, excelling in diverse real-world scenarios, delivering exceptional accuracy in singing lyrics recognition, and demonstrating robust generalization to Chinese dialects and English speech recognition.
By releasing model weights and inference code, we aim to contribute to the advancement of speech processing research. Future work will focus on further improving performance and expanding support for more languages and varied tasks.


\bibliographystyle{unsrt}
\bibliography{refs}


\end{document}